\documentclass[hidelinks]{article}
\usepackage{graphicx}
\usepackage{amssymb}
\usepackage{bbm}
\usepackage[utf8]{inputenc}
\usepackage[english]{babel}
\usepackage{amsmath}
\PassOptionsToPackage{hyphens}{url}\usepackage{hyperref}
\usepackage[toc,page]{appendix}

\usepackage[numbers, sort&compress]{natbib}
\setcitestyle{authoryear,open={(},close={)}} 

\usepackage[font={footnotesize,it}]{caption}

\usepackage{amsfonts}
\usepackage[left=.75in,top=.75in,right=.75in,bottom=.75in]{geometry}
\usepackage{setspace}
\usepackage{xcolor}

\usepackage{threeparttable}
\usepackage{url}
\usepackage[section]{placeins}

\title{Adaptive Selection of the Optimal Strategy to Improve Precision and Power in Randomized Trials}
\author{Laura B. Balzer, 
Erica Cai, 	Lucas Godoy Garraza, and Pracheta Amaranath}
\date{September 6, 2023}

\begin{document}

\maketitle

\begin{abstract}
Benkeser et al. demonstrate how adjustment for baseline covariates in randomized trials can meaningfully improve precision for a variety of outcome types.
Their findings build on a long history, starting in 1932 with R.A. Fisher and including more recent endorsements by the U.S. Food and Drug Administration and the European Medicines Agency. Here, we address an important practical consideration: \emph{how} to select the adjustment approach ---which variables and in which form--- to maximize precision, while maintaining Type-I error control. Balzer et al. previously proposed \emph{Adaptive Prespecification} within TMLE to flexibly and automatically select, from a prespecified set, the approach that maximizes empirical efficiency in small trials (N$<$40). To avoid overfitting with few randomized units, selection was previously limited to working generalized linear models, adjusting for a single covariate. Now, we tailor Adaptive Prespecification to trials with many randomized units. Using $V$-fold cross-validation and the estimated influence curve-squared as the loss function, we select from an expanded set of candidates, including modern machine learning methods adjusting for multiple covariates. As assessed in simulations exploring a variety of data generating processes, our approach maintains Type-I error control (under the null) and offers substantial gains in precision---equivalent to 20-43\% reductions in sample size for the same statistical power. When applied to real data from ACTG Study 175, we also see meaningful efficiency improvements overall and within  subgroups. \\
\end{abstract}

%


\maketitle


%

\section{Introduction}
\label{s:intro}

There is a long history of debate on whether and how to optimally adjust for baseline covariates to improve precision in randomized trials (e.g.,  \cite{Fisher1932, 
Tsiatis2008, Zhang2008, Europe2015, FDA_COVID}). Recently, for binary, ordinal, and time-to-event outcomes, \cite{Benkeser2020} defined several potential causal effects of interest and, for each, demonstrated the promise of covariate adjustment to improve our ability to make timely and precise inferences, without fear of bias due to regression model misspecification. In their simulation study, covariate adjustment led to substantial gains in efficiency, translating to 4-18\% reductions in sample size for the same statistical power when there was an effect, while maintaining good Type-I error control when there was no effect. 

However, \cite{Benkeser2020} only briefly discuss how to optimally select the adjustment covariates \emph{and} the ``working'' regression model to maximize precision for the effect of interest. In their Rejoinder, they  state, ``the variables should either be selected before the trial starts (selecting those that are most prognostic for the outcome based on prior data), or selected using the trial data based on a completely prespecified algorithm that aims to select the most prognostic variables'' \citep{Benkeser2020}. Their recommendation to use prior data assumes the existence of such data and consistency of relationships over time and space. Their recommendation for data-adaptive selection 
does not provide an specific algorithm.

The challenge of ``how'' 
is further highlighted in the corresponding commentaries. Specifically, \cite{ZhangZhang2021} emphasize that using a misspecified regression model for covariate adjustment can improve efficiency ``as long as the covariates are predictive of the outcomes''. This leaves the reader wondering about the potential detriments to efficiency and Type-I error control with forced adjustment for covariates that are, in fact, not  prognostic of the outcome.  \cite{ZhangZhang2021} 
call for further investigation into practical challenges, such as few independent units, stratified randomization, and covariate selection. 

Building on our previous work in Adaptive Prespecification (APS), we offer concrete solutions to these practical challenges \citep{Balzer2016DataAdapt}. Our approach data-adaptively selects, from a prespecified set, the covariates and the form of the working model to minimize the cross-validated variance estimate and, thereby, maximize the empirical efficiency. 
Our approach is applicable to asymptotically linear estimators with known influence curves and, thus, covers a large class of algorithms including those most commonly used for causal inference. Additionally, our work is applicable under a variety of trial designs: simple randomization, randomization within strata of baseline covariates, and randomization within matched pairs of units. Throughout we focus on a targeted minimum-loss based estimation (TMLE), which is summarized below  and detailed in Appendix A \citep{MarkBook}.
However, we emphasize APS can be applied to select the optimal, asymptotically linear estimator for a wide variety of causal effects.


\section{Methods}
\label{Methods}
In randomized trials, TMLE is a powerful approach to leverage baseline covariates for efficiency gains, while remaining robust to regression model misspecification (e.g., \cite{Moore2009, Rosenblum2010, Balzer2021twostage, Benitez2021}). For outcomes measured completely in a two-armed trial, 
the steps of TMLE 
are 
(1) Obtain an initial estimator of the “outcome regression”, defined as the conditional expectation of the outcome $Y$ given the treatment indicator $A$  and covariates $W$; 
(2) Obtain predicted outcomes under the treatment  $\hat{\mathbb{E}}(Y|A=1,W)$ and under the control  $\hat{\mathbb{E}}(Y|A=0,W)$;
(3) Target these outcome predictions using information in the ``propensity score'', defined as the conditional probability of receiving the treatment given the covariates $\mathbb{P}(A=1|W)$;
(4) Average the targeted predictions under the treatment $\hat{\mathbb{E}}^*(Y|A=1,W)$, and under the control $\hat{\mathbb{E}}^*(Y|A=0,W)$; 
and (5) Contrast on the scale of interest.

Under standard regularity conditions,  TMLE is asymptotically linear; therefore, the standardized estimator is asymptotically normal with mean zero and variance given by the variance of its influence curve (Appendix A). 
 If the initial estimator for the outcome regression $\mathbb{E}(Y|A,W)$ uses a ``working'' generalized linear model (GLM)  with an intercept and a main term for the treatment and if the propensity score is not estimated, then targeting (Step 3) can be skipped \citep{Rosenblum2010}. Further precision can be attained by using a prespecified, data-adaptive algorithm for initial estimation of the outcome regression and for ``collaborative'' estimation of the propensity score, as described next. 

\subsection{Adaptive Prespecification (APS)}

\cite{Rubin2008} proposed the principle of empirical efficiency maximization to optimize precision and, thus, power in randomized trials. To pick the ``best'' estimator, they propose using as loss function  the squared-efficient influence curve for the estimand of interest;  this corresponds to selecting the candidate TMLE with the smallest cross-validated variance estimate (\cite{MarkBook}; pp. 572--577). Building on this principle and motivated by the SEARCH community randomized trial ($N=32$), 
\cite{Balzer2016DataAdapt} proposed and implemented Adaptive Prespecification (APS) to select the optimal adjustment approach in trials with few randomized units.  We now generalize APS for use in trials with many randomized units. 

\begin{figure}
\begin{center}
\includegraphics[width=1\linewidth]{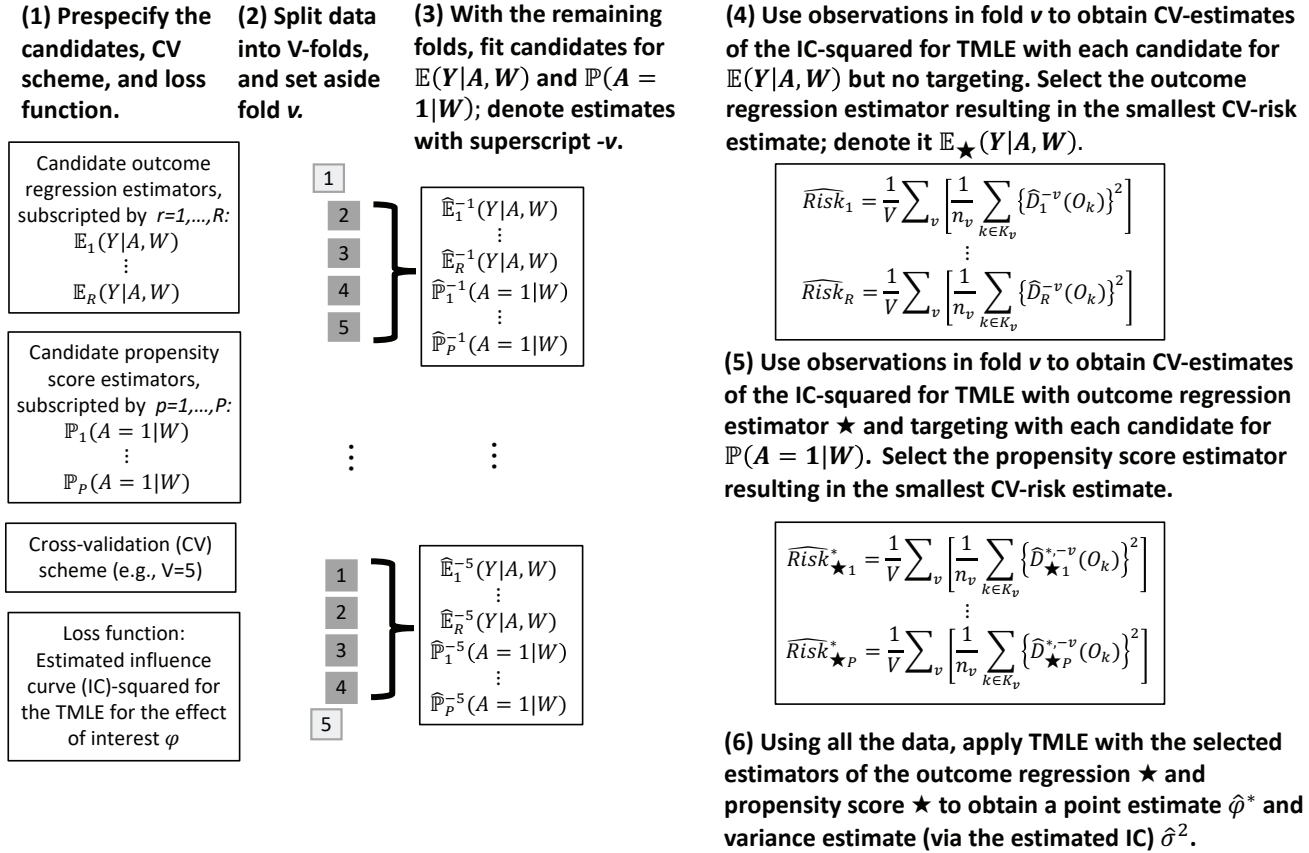}
\caption{Schematic of Adaptive Prespecification (APS) within TMLE to flexibly and automatically select, from a prespecified set, the adjustment approach that maximizes empirical efficiency for the effect of interest. For illustration, we show $R$ candidate outcome regression estimators $\mathbb{E}(Y|A,W)$, $P$ candidate propensity score estimators $\mathbb{P}(A=1|W)$, and $V=5$ fold cross-validation (CV). 
For simplicity, we show the process for first and last folds, and use ellipses to indicate an analogous process for the other folds. 
Let $K_v$ denote the set of indices for the observations in fold $v$ of size $|K_v|=n_v$.
For observation $k$ in validation set $v$, the CV-influence curve estimate for the TMLE using candidate outcome regression $r$ but no targeting is denoted $\hat{D}_r^{-v}(O_k)$ in Step 4, 
while the corresponding  CV-estimate of the influence curve for the TMLE using the selected outcome regression $\star$ and targeting with candidate propensity score estimator $p$ is denoted
$\hat{D}_{\star p}^{*,-v}(O_k)$
 (Appendix A). 
}
\label{f:schematic}
\end{center}
\end{figure}

Figure~\ref{f:schematic} provides a schematic of APS to select and implement the optimal TMLE (details in  Appendix A). 
First, we prespecify candidate estimators of the outcome regression and of the known propensity score.
Together they form the set of candidate TMLEs, which are consistent and locally efficient estimators.
As originally formulated for small trials, we limited candidate estimators of the outcome regression  to working GLMs with an intercept, a main term for the treatment $A$, and at most 1 adjustment variable. We also limited candidate estimators of the propensity score to working logistic regressions with an intercept and at most 1 adjustment variable. 
For the large-trial implementation, we now propose adjusting for multiple covariates in more flexible algorithms, such as penalized regression and multivariate adaptive regression splines (MARS). 
The unadjusted estimator must also be included as a candidate.
APS also requires us to prespecify a cross-validation scheme and a loss function to objectively measure performance. For small trials, we used leave-one-out cross-validation; for larger trials, we use V-fold cross-validation. 
As loss function, we use the estimated influence curve-squared for the TMLE of the effect of interest (Appendix A).

With these ingredients, we have fully prespecified and automated procedure to data-adaptively select the 
TMLE that maximizes empirical efficiency. Selection occurs in 2 steps. 
First, we select the candidate estimator of the outcome regression ---both the adjustment variable(s) and the functional form--- that minimizes the cross-validated variance estimate. Next, we select the candidate estimator of the propensity score ---both the adjustment variable(s) and the functional form--- that further minimizes the cross-validated variance estimate when used to target initial predictions from the previously selected outcome regression estimator. 
The two selected estimators form the optimal TMLE,
which is then fit using all the data. 

APS can be considered an extension of Collaborative-TMLE using a cross-validation selector (a.k.a., discrete Super Learner) to maximize precision in randomized trials.
In the small trial setting, substantial precision gains from APS have repeatedly been demonstrated in simulations and with real data \citep{Balzer2016DataAdapt,  Benitez2021}. For example, in the SEARCH study ($N=32$), we found that the variance of the unadjusted effect estimator was 4.6 times that of TMLE with the small-trial implementation of APS  \citep{Balzer2021twostage}. We now examine the performance of our proposed modification to APS for large trial settings using simulations as well as a real data application. 

\section{Simulation Studies}
\label{Sec:sims}

To address the persistent concerns about adjustment with non-linear models, highlighted by \cite{LaVange2021} among others, we explore data generating processes with higher order interactions and non-linear link functions. We also evaluate the performance with simple randomization and stratified randomization. Our focus is on estimating the sample effect; however, as previously discussed, our approach is 
applicable
to other asymptotically linear estimators with known influence curves, such as TMLE for the population average treatment effect (PATE) or the conditional average treatment effect (CATE; Appendix A).


We consider 5000 simulated trials, each with $N=500$ participants. For each participant,
we generate 5 measured covariates $\{W_1,\ldots, W_5\}$ from a standard normal distribution, 2 unmeasured covariates $\{U_{1},U_{2}\}$ from a standard uniform distribution, and the binary counterfactual outcomes $Y(a)$ in 3 settings of 
varying complexity: 
\begin{enumerate}
    \item ``Linear'': $Y(a) =\mathbbm{1}\big\{U_{1} < logit^{-1}(a+W_1-W_2+W_3-W_4 +W_5 - 2aW_1 + U_{2}) \big\}$ 
 \item ``Interactive'':  $Y(a) = \mathbbm{1}\big\{U_{1} < logit^{-1}(a+W_1+W_2+W_3+W_4+W_5+aW_1 +aW_2W_4 + aW_3 + aW_5U_{2} + U_2) \big\}$ 
\item ``Polynomial'':  $Y(a) = \mathbbm{1}\big\{U_{1} < logit^{-1}(a+W_1+W_2+W_3+W_4+W_5 - W_1W_3+2W_1W_3W_4-W_4(1-W_1) + U_{2})\big\}$
\end{enumerate}
We additionally consider a ``Treatment only'' scenario where none of the measured covariates influences the outcome: $Y(a) = \mathbbm{1}\big\{U_{1} < logit^{-1}(0.1a + 2U_{2})\big\}$.
Using these counterfactual outcomes, we calculate the true value of 
the sample risk ratio (RR)=$1/N \sum_i Y_i(1) \div 1/N \sum_i Y_i(0)$.
 As detailed in Appendix B, we also consider a continuous outcome, generated under 4 analogous settings. For the continuous endpoint, we focus on  the sample average treatment effect (SATE)=$1/N \sum_i Y_i(1) - 1/N \sum_i Y_i(0)$. 
In each setting, we generate the observed treatment $A$ using simple randomization and randomization within strata defined by $\mathbbm{1}(W_1>0)$. Finally, we set the observed outcome $Y$ equal to the counterfactual outcome $Y(a)$ when the observed treatment $A=a$. 

We compare the unadjusted estimator, fixed adjustment for $W_1$ in the outcome regression, TMLE with APS tailored for small trials, and TMLE with our novel modification of APS for larger trials. In the small-trial APS, we limit the candidate estimator  to working GLMs with 1 adjustment covariate selected from $\{W_1, W_2, W_3, W_4, W_5, \emptyset\}$. In the large-trial APS, we select from working GLMs adjusting for 1  covariate, main terms GLM adjusting for all covariates, stepwise regression, stepwise regression with all possible pairwise interactions, LASSO, MARS, and the unadjusted estimator. Both versions of APS use 5-fold cross-validation. 
Performance criteria include 95\% confidence interval coverage, attained power, Type-I error (under the null), bias, variance, and mean squared error (MSE). Following \cite{Benkeser2020}, we  provide the relative efficiency, calculated as the MSE of a covariate-adjusted estimator divided by the MSE of the unadjusted effect estimator, and provide an estimate the potential savings in sample size, calculated as 1 minus the relative efficiency. 

\begin{figure}
\begin{center}
\includegraphics[width=.75\linewidth]{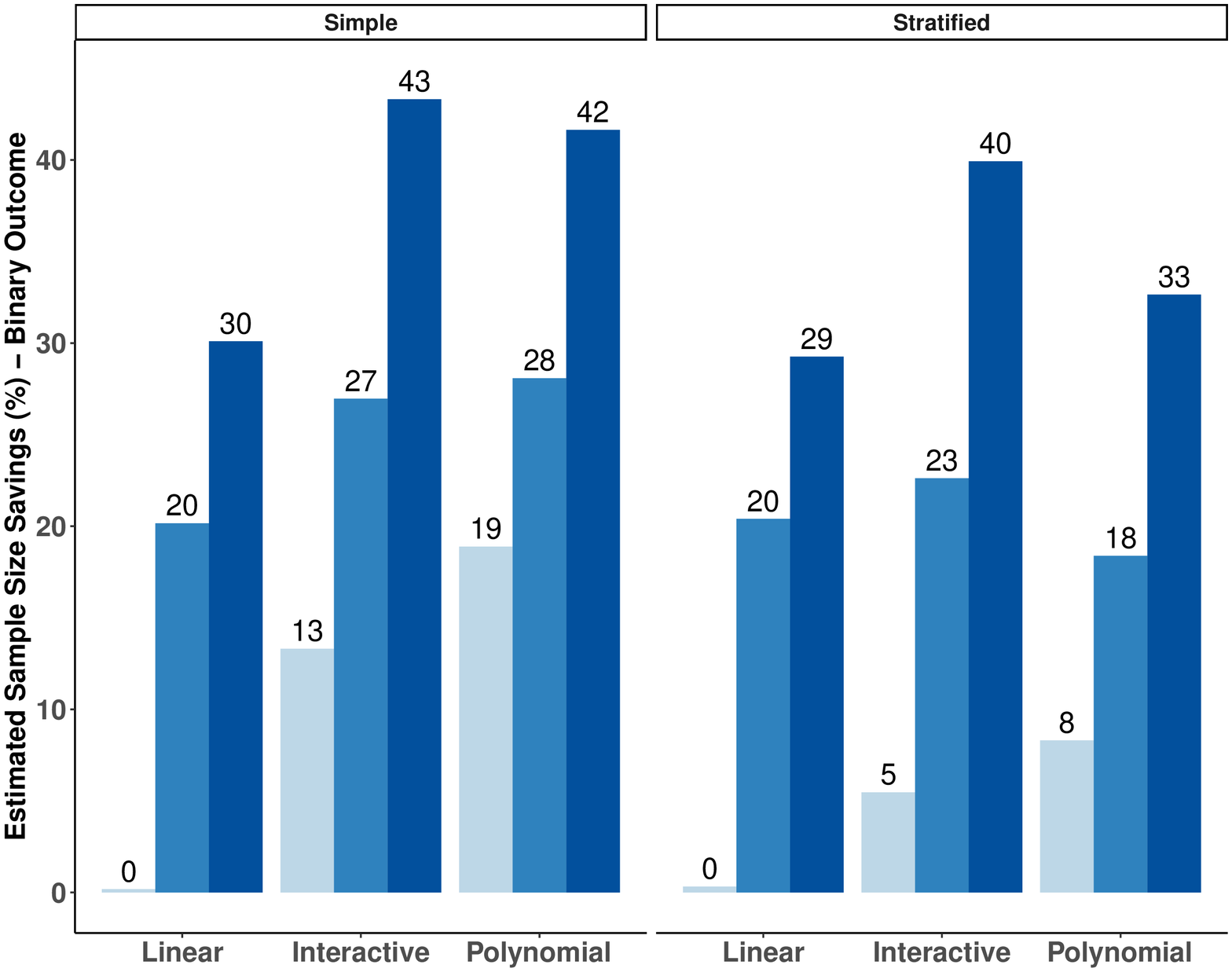}
\includegraphics[width=.75\linewidth]{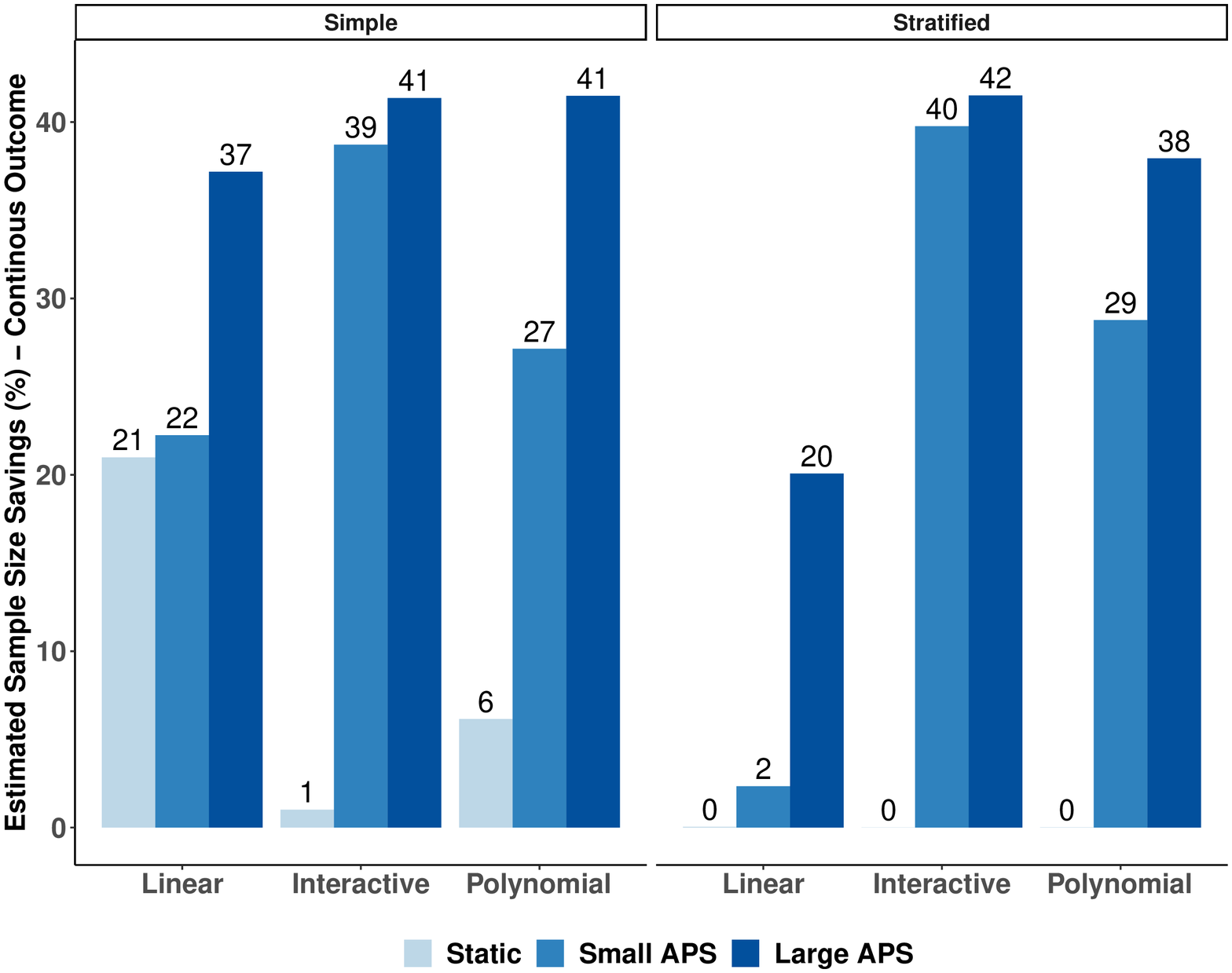}
\caption{Across 5000 simulated trials with a  binary outcome (top) and with a continuous  outcome (bottom), the estimated savings in sample size (in \%), as compared to the unadjusted estimator, when using  forced adjustment for $W_1$ in the outcome regression (``Static''), TMLE with the small-trial implementation of Adaptive Prespecification (``Small APS''), and TMLE with the large-trial implementation (``Large APS'') across the 3 data generating processes with prognostic covariates and with simple versus stratified randomization.}
\label{f:savings}
\end{center}
\end{figure}

%
For the binary outcome, TMLE using the large-trial APS  substantially improved power for the risk ratio in all scenarios (Table~\ref{Table:BinEffect}). Absolute gains in power as compared to the unadjusted approach ranged from 5\% to 11\%. The relative efficiency ranged from 0.57 to 0.71. This roughly translates to 29-43\% savings in sample size from using TMLE, instead of the unadjusted effect estimator (Figure~\ref{f:savings}). 
 The  gains from TMLE with the small-trial APS were less extreme, but still notable (relative efficiency: 0.72-0.82).
Importantly, all estimators maintained nominal-to-conservative confidence interval coverage, as expected when estimating sample effects (Appendix A). 

 \begin{table}
\caption{Estimator performance with the binary outcome where there is an effect and with a sample size of $N=500$.}
\label{Table:BinEffect}
\begin{threeparttable}[t]
\centering
\begin{tabular}{l l l llllll}
  \hline
DGP & Design & Estimator & Cover. & Power & MSE & Bias & Var. & Rel.Eff. \\ 
  \hline
Linear & Simple & Unadjusted & 0.976 & 0.929 & 0.005 & 0.002 & 0.007 & 1.000 \\ 
   &  & Static & 0.977 & 0.929 & 0.005 & 0.002 & 0.007 & 0.998 \\ 
   &  & Small APS & 0.982 & 0.950 & 0.004 & -0.006 & 0.006 & 0.798 \\ 
   &  & Large APS & 0.971 & 0.984 & 0.004 & 0.002 & 0.006 & 0.699 \\ 
   & Stratified & Unadjusted & 0.980 & 0.935 & 0.005 & 0.003 & 0.007 & 1.000 \\ 
   &  & Static & 0.979 & 0.935 & 0.005 & 0.003 & 0.007 & 0.997 \\ 
   &  & Small APS & 0.984 & 0.952 & 0.004 & -0.005 & 0.006 & 0.796 \\ 
   &  & Large APS & 0.974 & 0.984 & 0.004 & 0.003 & 0.006 & 0.707 \\ 
  Interactive & Simple & Unadjusted & 0.965 & 0.211 & 0.003 & 0.003 & 0.003 & 1.000 \\ 
   &  & Static & 0.964 & 0.228 & 0.002 & 0.003 & 0.003 & 0.867 \\ 
   &  & Small APS & 0.975 & 0.212 & 0.002 & 0.001 & 0.002 & 0.730 \\ 
   &  & Large APS & 0.965 & 0.324 & 0.002 & 0.003 & 0.002 & 0.567 \\ 
   & Stratified & Unadjusted & 0.970 & 0.184 & 0.003 & 0.001 & 0.003 & 1.000 \\ 
   &  & Static & 0.965 & 0.216 & 0.002 & 0.001 & 0.003 & 0.945 \\ 
   &  & Small APS & 0.980 & 0.190 & 0.002 & -0.001 & 0.002 & 0.774 \\ 
   &  & Large APS & 0.966 & 0.297 & 0.002 & 0.001 & 0.002 & 0.601 \\ 
  Polynomial & Simple & Unadjusted & 0.963 & 0.865 & 0.004 & 0.004 & 0.005 & 1.000 \\ 
   &  & Static & 0.967 & 0.916 & 0.004 & 0.002 & 0.004 & 0.811 \\ 
   &  & Small APS & 0.970 & 0.914 & 0.003 & -0.004 & 0.004 & 0.719 \\ 
   &  & Large APS & 0.969 & 0.969 & 0.003 & 0.001 & 0.003 & 0.584 \\ 
   & Stratified & Unadjusted & 0.977 & 0.868 & 0.004 & 0.002 & 0.004 & 1.000 \\ 
   &  & Static & 0.966 & 0.912 & 0.003 & 0.002 & 0.004 & 0.917 \\ 
   &  & Small APS & 0.974 & 0.908 & 0.003 & -0.004 & 0.004 & 0.816 \\ 
   &  & Large APS & 0.972 & 0.972 & 0.003 & 0.001 & 0.003 & 0.673 \\ 
   \hline
\end{tabular}
\begin{tablenotes}
\footnotesize 
\item[] ``DGP'' denotes the data generating process; ``Cover.'' denotes the 95\% confidence interval coverage; ``Power'' denotes the proportion of times the true null hypothesis was rejected; ``MSE'' denotes mean squared error; ``Var.'' denotes the variance of the point estimates, and ``Rel.Eff.'' denotes relative efficiency, approximated by the ratio of the MSE of a given estimator to that of the unadjusted estimator.  
The average value of the sample risk ratio is 1.25 in the Linear setting, 1.06 in the Interactive setting, and 1.19 in the Polynomial setting.
``Static'' refers to forced adjustment for $W_1$ in the outcome regression, ``Small APS'' to TMLE with the small-trial implementation of Adaptive Prespecification (APS), and ``Large APS'' to TMLE with the large-trial implementation of APS.
\end{tablenotes}
\end{threeparttable}
\end{table}

Estimator performance with a continuous outcome and targeting SATE was similar (Table~\ref{Table:ContEffect}). In all scenarios, TMLE with the large-trial APS achieved the highest power with absolute gains of 18-23\% compared to the unadjusted estimator. Its relative efficiency was 0.58-0.80, roughly translating to 20-42\% savings in sample size (Figure~\ref{f:savings}).  As before, the  gains from TMLE with the small-trial APS were less extreme, but still notable (relative efficiency: 0.60-0.98). 
Again, the 95\% confidence interval coverage of the adaptive estimators was comparable to that of the unadjusted estimator.

\begin{table}
\caption{Estimator performance with the continuous outcome where there is an effect and with a sample size of $N=500$.}
\label{Table:ContEffect}
\begin{threeparttable}[t]
\centering
\begin{tabular}{l l l llllll}
  \hline
DGP & Design & Estimator & Cover. & Power & MSE & Bias & Var. & Rel.Eff. \\ 
  \hline
Linear & Simple & Unadjusted & 0.953 & 0.591 & 0.008 & -0.001 & 0.008 & 1.000 \\ 
   &  & Static & 0.948 & 0.692 & 0.006 & -0.002 & 0.006 & 0.790 \\ 
   &  & Small APS & 0.950 & 0.689 & 0.006 & -0.002 & 0.006 & 0.778 \\ 
   &  & Large APS & 0.950 & 0.794 & 0.005 & -0.001 & 0.005 & 0.628 \\ 
   & Stratified & Unadjusted & 0.971 & 0.617 & 0.006 & 0.001 & 0.006 & 1.000 \\ 
   &  & Static & 0.945 & 0.705 & 0.006 & 0.001 & 0.006 & 1.000 \\ 
   &  & Small APS & 0.947 & 0.706 & 0.006 & 0.000 & 0.006 & 0.977 \\ 
   &  & Large APS & 0.944 & 0.799 & 0.005 & 0.001 & 0.005 & 0.799 \\ 
  Interactive & Simple & Unadjusted & 0.948 & 0.389 & 0.011 & -0.001 & 0.011 & 1.000 \\ 
   &  & Static & 0.946 & 0.399 & 0.011 & -0.001 & 0.011 & 0.990 \\ 
   &  & Small APS & 0.948 & 0.578 & 0.007 & -0.001 & 0.007 & 0.613 \\ 
   &  & Large APS & 0.946 & 0.602 & 0.007 & -0.001 & 0.007 & 0.586 \\ 
   & Stratified & Unadjusted & 0.942 & 0.402 & 0.012 & 0.003 & 0.012 & 1.000 \\ 
   &  & Static & 0.939 & 0.409 & 0.012 & 0.003 & 0.012 & 1.000 \\ 
   &  & Small APS & 0.946 & 0.586 & 0.007 & 0.001 & 0.007 & 0.602 \\ 
   &  & Large APS & 0.943 & 0.617 & 0.007 & 0.001 & 0.007 & 0.585 \\ 
  Polynomial & Simple & Unadjusted & 0.948 & 0.489 & 0.008 & -0.001 & 0.008 & 1.000 \\ 
   &  & Static & 0.943 & 0.518 & 0.007 & -0.001 & 0.007 & 0.938 \\ 
   &  & Small APS & 0.953 & 0.613 & 0.006 & -0.002 & 0.006 & 0.729 \\ 
   &  & Large APS & 0.948 & 0.724 & 0.004 & -0.001 & 0.004 & 0.585 \\ 
   & Stratified & Unadjusted & 0.948 & 0.508 & 0.007 & 0.002 & 0.007 & 1.000 \\ 
   &  & Static & 0.939 & 0.536 & 0.007 & 0.002 & 0.007 & 1.000 \\ 
   &  & Small APS & 0.957 & 0.631 & 0.005 & 0.001 & 0.005 & 0.712 \\ 
   &  & Large APS & 0.942 & 0.735 & 0.005 & 0.001 & 0.005 & 0.621 \\ 
   \hline
\end{tabular}
\begin{tablenotes}
\footnotesize 
\item[] ``DGP'' denotes the data generating process; ``Cover.'' denotes the 95\% confidence interval coverage; ``Power'' denotes the proportion of times the true null hypothesis was rejected; ``MSE'' denotes mean squared error; ``Var.'' denotes the variance of the point estimates, and ``Rel.Eff.'' denotes relative efficiency, approximated by the ratio of the MSE of a given estimator to that of the unadjusted estimator.  The average value of the sample average treatment effect (SATE) is 0.195 in the Linear setting, 0.180 in the Interactive setting, and 0.170 in the Polynomial setting. ``Static'' refers to forced adjustment for $W_1$ in the outcome regression, ``Small APS'' to TMLE with the small-trial implementation of Adaptive Prespecification (APS), and ``Large APS'' to TMLE with the large-trial implementation of APS.
\end{tablenotes}
\end{threeparttable}
\end{table}

Additional results are available in Appendix B. The selection of estimators for the outcome regression and propensity score varied by setting, highlighting our approach's ability to respond to the data generating process (Tables 4-5). Importantly, for both outcome types, the precision gains from TMLE were achieved without sacrificing Type-I error, even in the ``Treatment only'' setting where there were no prognostic covariates (Tables 6-7).

\section{Real data application: ACTG Study 175}

ACTG Study 175 evaluated the effect of monotherapy versus combination therapy on health outcomes among persons with HIV 
\citep{ACTG175}. For demonstration, we examine the effect of the antiretroviral therapy (ART) regimen only containing zidovudine ($A=0$) versus on alternative regimen ($A=1)$  on the difference in the average CD4 count at 20 weeks (continuous outcome) and  the relative risk of the 20-week CD4 count$>$350c/mm$^3$ (binary outcome). We use the unadjusted estimator, fixed adjustment for age and gender, TMLE with the small-trial APS, and TMLE with the large-trial APS. 
Within APS, we considered 16 candidate adjustment variables, including demographics and ART history (Table 8). 

\begin{table}
\caption{Comparative results using real data from ACTG Study 175 to estimate the effect on difference in the average CD4 count at 20 weeks (continuous outcome) and on the relative risk of 20-week CD4 count $>$350c/mm$^3$ (binary outcome). Additional details and results are given in Appendix C.}
\label{Table:app}
\centering
\begin{threeparttable}[t]
\begin{tabular}{lllllll}
  \hline
Outcome  & Estimator & Effect (95\%CI) & Rel.Var. &Out.Reg. & PScore  & Type-I \\ 
  \hline
Continuous & Unadjusted & 46.4 (33.0, 59.7) & 1.000 & Unadj. & Unadj. & 4.8\% \\ 
   & Static & 46.8 (33.5, 60.0) & 0.991 & Fixed & Fixed & 4.8\% \\ 
   & Small APS & 48.5 (38.0, 59.0) & 0.617 & GLM & GLM & 5.2\% \\ 
   & Large APS & 47.8 (38.0, 57.6) & 0.542 & MARS & GLM & 5.3\% \\ 
  Binary & Unadjusted & 1.23 (1.10, 1.37) & 1.000 & Unadj. & Unadj. & 4.7\% \\ 
   & Static & 1.23 (1.11, 1.37) & 1.001 & Fixed & Fixed & 4.5\% \\ 
   & Small APS & 1.26 (1.15, 1.38) & 0.702 & GLM & GLM & 5.2\% \\ 
   & Large APS & 1.26 (1.15, 1.37) & 0.672 & LASSO & GLM & 5.2\% \\ 
   \hline
\end{tabular}
\begin{tablenotes}
\footnotesize 
\item[]  
``Static'' refers to forced adjustment for age in the outcome regression and gender in the propensity score, ``Small APS'' to TMLE with the small-trial implementation of Adaptive Prespecification (APS), selecting from working GLMs adjusting for at most 1 covariate,  and ``Large APS'' to TMLE with the large-trial implementation of APS, selecting from the small-trial algorithms, main terms, stepwise regression, LASSO, MARS, and MARS after correlation-based screening.
``Rel.Var.'' is the estimated variance of a given approach divided by the estimated variance of the unadjusted approach. 
``Out.Reg.'' is the selected approach for estimation of the outcome regression, and ``PScore'' is the selected approach for estimation of the known propensity score. ``GLM'' refers to a working GLM adjusted for at most 1 covariate.
``Type-I'' is the estimated Type-I error rate, evaluated with treatment-blind simulations, which permute the treatment indicator $A$, implement each estimator, and repeat 5000 times.  
\end{tablenotes}
\end{threeparttable}
\end{table}

The results are summarized in Table~\ref{Table:app} with further details in Appendix C. As expected, the point estimates were similar across approaches, but application of APS offered notable precision gains. The estimated variance of TMLE with the large-trial APS divided by that of the unadjusted approach was 0.54 and 0.67 for the continuous and binary outcome, respectively. Assuming negligible bias, this would roughly translate into needing 46\% and 33\% fewer participants with our approach. Importantly, Type-I error control, evaluated through treatment-blind simulations, 
 was maintained at the nominal rate of 5\%. As expected, the optimal TMLE varied by the target of inference and sample size (Tables 9-10). For smaller subgroups of older  and younger women, 
 there were notable gains in efficiency from estimation of the 
 propensity score, but no difference between the TMLEs using the large- vs. small-trial APS. Here, both APS implementations selected a working GLM adjusting for 1 covariate when estimating the outcome regression and when estimating  the  propensity score. In contrast, overall and for larger subgroups of older and younger men, TMLE with the large-trial APS offered notable precision gains over the small-trial implementation, but there were minimal precision improvements from propensity score estimation. 

\section{Discussion}
\label{s:discuss}

The U.S. Food and Drug Administration and the European Medicines Agency endorse adjustment for baseline covariates to improve precision and, thereby, power in randomized trials \citep{Europe2015, FDA_COVID}. Nonetheless, explicit guidance on how to optimally select and incorporate adjustment variables has been lacking. Indeed, the challenges in practical implementation were discussed in  \cite{Benkeser2020} and the accompanying commentaries.
For trials with limited numbers of randomized units ($N<40)$, \cite{Balzer2016DataAdapt} addressed this gap with \emph{Adaptive Prespecification} (APS), which selects the adjustment strategy maximizing the empirical efficiency. 
Here, for trials with many randomized units, we extended APS to include machine learning algorithms (e.g., LASSO and MARS) adjusting for multiple covariates. 
Our simulations demonstrated improved precision and power, translating to 18-43\% potential savings in sample size, while maintaining Type-I error. These gains were seen across  a variety of data generating processes, for both binary and continuous outcomes, and for both absolute and relative effects. Our real data application also demonstrated precision gains and highlighted how selection of the optimal TMLE was responsive to the subgroup.

 Our approach offers several advantages over other model-robust, covariate-adjusted estimators. First, it is estimand-aligned; we can estimate user-specified effects on any scale (e.g., difference, ratio), for any inferential target (e.g., sample, conditional, or population effect), for several study designs (e.g., simple, stratified, matched), and for a variety of outcome types (e.g., binary, continuous). Second, our approach is fully prespecified, while remaining data-adaptive. Practically, we can prespecify several candidate estimators of the outcome regression, and let the algorithm pick the best approach, where ``best'' means maximizing empirical efficiency. These candidates can include user-specified GLMs, including known or suspected interactions, as well as modern advances in machine learning. Third, our approach incorporates collaborative estimation of the known propensity score for additional gains in precision. Thereby, we only estimate  $\mathbb{P}(A=1|W)$  if it improves the empirical efficiency; otherwise, we treat the propensity score as known and only adjust in the outcome regression. Collaborative estimation of the propensity score does come at the cost of a more complicated algorithm. However, it can meaningfully improve precision, especially in small trials \citep{Balzer2016DataAdapt} or smaller subgroups (Tables 9-10), and computing code is readily available. Finally, if we are in the unfortunate scenario where adjustment does not improve efficiency, the algorithm will default to the unadjusted effect estimator. Thus, we are protected from forced adjustment at the detriment of precision or Type-I error control.

 We have the following recommendations when implementing APS.  First, increase the number of cross-validation folds as the number of randomized units decreases. Second, as candidates, consider a diverse set of asymptotically linear estimators. To prevent forced adjustment when harmful to precision, always include the unadjusted estimator as a candidate. As shown here, including candidates that flexibly adjust for multiple covariates, while satisfying the usual regularity conditions, can lead to substantial savings in sample size (Figure~\ref{f:savings}).
If considering more aggressive algorithms (e.g., random forest) that do not readily satisfy the conditions for asymptotic linearity, additional sample-splitting is recommended. 
 APS naturally generates a cross-validated variance estimate, which can be used if there are concerns about overfitting. 
 To guide development of the Statistical Analysis Plan,  we strongly recommend conducting a simulation study, reflecting the real data application, to facilitate
  prespecification of the candidate estimators, the cross-validation scheme, and the variance estimator based on objective criteria (e.g., relative efficiency and Type-I error control).
 
There are several limitations to our presentation. First, we focused on using APS to chose between candidate TMLEs; however, the procedure is applicable to other asymptotically linear estimators with known influence curves. This includes more traditional estimators, such as the Cox model for time-to-event outcomes. We plan to apply APS to select the optimal approach from a variety of doubly robust candidates, 
such as TMLE, augmented inverse probability weighting, and double/debiased machine learning. Second, we focused on trials were outcomes are measured  completely; our work is immediately applicable to settings where censoring or missingness is completely at random. 
If, instead,  censoring or missingness is random conditional on a $X$, a subset of the full covariate set $W$, APS 
should also be applicable with the following modification: all candidates must adjust for $X$ and may consider additional adjustment for the remaining covariates.
Further investigation is warranted. (We refer the reader to \cite{Balzer2021twostage} for application of APS in cluster randomized trials with missing or censored outcomes.)
Third, our approach is applicable to trials with simple randomization, stratified randomization, and randomization within matched pairs; further investigation is needed for settings with sequential randomization.
Finally, as currently implemented, APS uses a cross-validation selector (a.k.a., discrete Super Learner) to choose the single best estimator of the outcome regression combined with the single best estimator of the propensity score. We are working to extend APS to select the optimal convex combination of candidate estimators. %
Nonetheless, we believe TMLE with APS, as currently implemented, is a powerful and under-utilized tool for optimal covariate adjustment in randomized trials.

\section*{Acknowledgements}
This work was supported, in part, by the National Institutes of Health 
(NIH; U01AI150510, 
2R01AI074345-10A1) and  DARPA (HR001120C0031). 
We  thank Drs. Alan Hubbard, Maya Petersen, and Mark van der Laan for their  feedback. Any findings and recommendations are the authors and do not necessarily reflect the views of the NIH or the US Air Force.
\vspace*{-8pt}

\section*{Data Availability Statement}

The data that support the findings of this paper are openly available in GitHub at \url{https://github.com/LauraBalzer/AdaptivePrespec}. Computing code, including a vignette with worked examples, is also available via this GitHub.

\section*{Appendix A. More on TMLE and Adaptive Pre-specification (APS)}


Recall $W$ denotes the baseline covariates; $A$ is an indicator of being randomized to the intervention, and $Y$ is the outcome. Let $Y_i(a)$ denote the counterfactual outcome  for randomized unit (i.e.,  participant)  $i=\{1, \ldots,N\}$ under treatment-level $A=a$. 
When defining the causal effects, we take contrasts of ``treatment-specific means'', defined  at the population-level $\psi^p(a)=\mathbb{E}\big[Y(a)\big]$, conditional on covariates
 $\psi^c(a) =\frac{1}{N} \sum_{i=1}^N \mathbb{E}\big[Y_i(a)\big |W_i \big]$, or for the study sample $\psi^s(a)= \frac{1}{N} \sum_{i=1}^N Y_i(a)$ \citep{Neyman1923, Imbens2004, Imai2008, Balzer2015Adaptive, Balzer2016SATE}. For example, the population average treatment effect (PATE) is $\psi^p(1)-\psi^p(0)$; the conditional average treatment effect (CATE) is $\psi^c(1)-\psi^c(0)$, and  the sample average treatment effect (SATE) is $\psi^s(1)-\psi^s(0)$. Of course, these treatment-specific means can also be contrasted on the relative or odds ratio scale. 

\subsection*{A.1 Step-by-step implementation of TMLE in randomized trials}

For a review of TMLE and its relation to other effect estimators in randomized trials,  we refer to \cite{Colantuoni2015}.
For demonstration, we first focus on a binary outcome $Y\in \{0,1\}$. In this setting and as described by \cite{Benkeser2020}, we can adjust for covariates to improve precision when estimating the population risk difference (RD)=$\psi^p(1)-\psi^p(0)$ or risk ratio (RR)=$\psi^p(1)\div\psi^p(0)$, 
as follows \citep{Scharfstein1999,Moore2009,Rosenblum2010}:
\begin{enumerate}
\item Use a working logistic regression model to estimate the conditional probability of the outcome, given the intervention indicator and selected covariates:  $\mathbb{P}(Y=1|A,W)$.
\item Using the fit from Step \#1, predict the outcome under the intervention $\hat{\mathbb{P}}(Y=1|A=1,W_i)$ and under the control $\hat{\mathbb{P}}(Y=1|A=0,W_i)$  for $i=\{1,\ldots,N\}$.
\item Contrast the average arm-specific predictions to estimate the risk difference (RD) or risk ratio (RR), respectively: 

$$\hat{RD}  = \frac{1}{N}\sum_{i=1}^N  \hat{\mathbb{P}}(Y=1|A=1,W_i) - \frac{1}{N}\sum_{i=1}^N  \hat{\mathbb{P}}(Y=1|A=0,W_i)$$ 
$$\hat{RR} =\frac{1}{N}\sum_{i=1}^N  \hat{\mathbb{P}}(Y=1|A=1,W_i) \div \frac{1}{N}\sum_{i=1}^N \hat{\mathbb{P}}(Y=1|A=0,W_i) $$
\end{enumerate}
The same implementation is used for point estimation of conditional effects defined as contrasts of $\psi^c(a)$ 
and for sample effects defined as contrasts of $\psi^s(a)$
 \citep{Balzer2015Adaptive, Balzer2016SATE}.

Importantly, the working  regression model in Step 1 is robust to misspecification. It can include interactions and other user-specified terms. 
Furthermore, the number of parameters (i.e., coefficients) to be fit in Step \#1 would be equivalent to if we had, instead, used a linear working model. An important benefit of using logistic regression over linear regression is that the logit-link guarantees the bounds on the binary outcome will be respected \citep{Gruber2010}; this can be particularly important under data sparsity (e.g., due to rare outcomes). That said, when using Adaptive Prespecification (APS),  we can consider candidate outcome regression estimators using both linear and logit-links, and let the data decide which candidate maximizes empirical efficiency, as detailed below.

Doubly robust methods incorporate information in the propensity score when solving the efficient score equation. In TMLE this is done during the targeting step. In a randomized trial, the propensity score $\mathbb{P}(A=1|W)$ is known and does not need to be estimated. For example, in a two-armed trial with equal allocation probability: $\mathbb{P}(A=1|W)=\mathbb{P}(A=1)=0.5$. When the propensity score is not estimated, TMLE reduces to the above approach \citep{Rosenblum2010} and, therefore, requires estimation of the same number of regression parameters. 
However, collaborative estimation of the known propensity score has been shown repeatedly to increase efficiency (e.g., \cite{MarkRobins,MarkBook, Balzer2016DataAdapt}).

For a binary outcome or, more generally, a bounded continuous outcome, we now present the full TMLE algorithm, including estimation of the propensity score and subsequent targeting. Steps 1-2 are the same as before. We use $^*$ to denote targeted estimates.
\begin{enumerate}
\item Use a working logistic regression model to estimate the conditional expectation of the outcome, given the intervention indicator and selected covariates:  $\mathbb{E}(Y|A,W)$.
\item Using the fit from Step \#1, predict the outcome under the intervention $\hat{\mathbb{E}}(Y|A=1,W_i)$ and under the control $\hat{\mathbb{E}}(Y|A=0,W_i)$  for $i=\{1,\ldots,N\}$.
\item Target the initial outcome predictions $\hat{\mathbb{E}}(Y|A,W)$ using information in the estimated propensity score $\hat{\mathbb{P}}(A=1|W)$. The targeting approach is driven by the efficient influence curve and, thus, depends on the statistical estimand of interest \citep{MarkBook}.  One approach to simultaneously target effects defined on the difference, ratio, and odds ratio scales is to  use a two-dimensional ``clever covariate'' \citep{Moore2009}:
\begin{enumerate}
    \item Use a working logistic regression model to estimate the conditional probability of the intervention given the selected covariates: $\mathbb{P}(A=1|W)$.
    \item  Using the propensity score fit from the previous step, calculate the clever covariates  $\hat{H}(1,W_i) = \frac{\mathbbm{1}(A_i=1)}{\hat{\mathbb{P}}(A=1|W_i)}$ and $\hat{H}(0,W_i) = \frac{\mathbbm{1}(A_i=0)}{\hat{\mathbb{P}}(A=0|W_i)}$ for $i=\{1,\ldots, N\}$
    \item 
Estimate the fluctuation coefficients $\epsilon_1$ and $\epsilon_0$ in the following working logistic regression model:
\[logit[\mathbb{E}^*(Y|A,W)]= logit[\mathbb{E}(Y|A,W)] + \epsilon_1 \hat{H}(1,W) +  \epsilon_0\hat{H}(0,W)\] Note there is no intercept and the coefficient for the logit of the initial estimates is set to 1. In practice, we run logistic regression of the observed outcome $Y$ on the the clever covariates $\hat{H}(1,W)$ and $\hat{H}(0,W)$ with the logit of the initial estimates $\mathbb{E}(Y|A,W)$ as offset.
Using maximum likelihood estimation corresponds to solving the relevant component of the efficient score equation (a.k.a., the efficient influence curve equation) \citep{MarkBook}. Denote the resulting estimates as $\hat{\epsilon}_1$ and $\hat{\epsilon}_0$.
    \item For $i=\{1,\ldots,N\}$, obtain targeted outcome predictions under the intervention and control, calculated as  
    \[ \hat{\mathbb{E}}^*(Y|A=1,W_i) = logit^{-1}\left[ logit\{ \hat{\mathbb{E}}(Y|A=1,W_i)\} + \frac{\hat{\epsilon}_1}{\hat{\mathbb{P}}(A=1 |W_i)}  \right] \]
      \[ \hat{\mathbb{E}}^*(Y|A=0,W_i) = logit^{-1}\left[ logit\{ \hat{\mathbb{E}}(Y|A=0,W_i)\} + \frac{\hat{\epsilon}_0}{\hat{\mathbb{P}}(A=0 |W_i)}  \right] \]
\end{enumerate}
  
\item Contrast the average arm-specific, targeted predictions to estimate the risk difference or risk ratio, respectively: 

$$\hat{RD}^*  = \frac{1}{N}\sum_{i=1}^N  \hat{\mathbb{E}}^*(Y|A=1,W_i) - \frac{1}{N}\sum_{i=1}^N  \hat{\mathbb{E}}^*(Y|A=0,W_i)$$ 
$$\hat{RR}^* =\frac{1}{N}\sum_{i=1}^N  \hat{\mathbb{E}}^*(Y|A=1,W_i) \div \frac{1}{N}\sum_{i=1}^N \hat{\mathbb{E}}^*(Y|A=0,W_i) $$
\end{enumerate}
If the effect  on the difference scale is of primary interest, the targeting step can be simplified by using a one-dimensional clever covariate: $\hat{H}(A,W) = \frac{\mathbbm{1}(A_i=1)}{\hat{\mathbb{P}}(A=1|W_i)} - 
\frac{\mathbbm{1}(A_i=0)}{\hat{\mathbb{P}}(A=0|W_i)}$. 
We could even define candidate TMLEs using alternative targeting approaches and select the optimal using APS. 
In our extension of APS for trials with many randomized units, we consider more flexible candidate estimators of the outcome regression in Step \#1 and the propensity score in Step \#3.

\subsection*{A.2. Statistical inference \& influence curves (functions)}

Under standard regularity conditions, which are weak in randomized trials, TMLE is an asymptotically linear estimator of the population, conditional, and sample effects \citep{Moore2009, Rosenblum2010, MarkBook, Balzer2015Adaptive, Balzer2016SATE}. 
 Therefore, the estimator minus the estimand can be written as an empirical mean of its influence curve plus a remainder term going to 0 in probability. 
For example, for the TMLE $\hat{\psi}^{p*}(a)$ of the population parameter $\psi^p(a) =\mathbb{E}\big[Y(a)\big]$, we have
\[
\hat{\psi}^{p*}(a) - \psi^p(a) = \frac{1}{N}\sum_{i=1}^N D^{p*}(a,O_i) + o_P(1/\sqrt{N})
\]
where $D^{p*}(a,O_i)$ is the influence curve for observation $O_i$. (The form of the influence curve is given explicitly below.)
Asymptotically linear estimators enjoy properties following from the Central Limit Theorem. 
In particular, the standardized estimator is normally distributed in the large-data limit with variance given by the variance of its influence curve. In randomized trials, TMLE is locally efficient in that it will have the lowest possible asymptotic variance if the working model for the outcome regression is correctly specified; then its influence curve will equal the efficient influence curve. 

The variance of asymptotically linear estimators is conveniently estimated by the sample variance of the estimated influence curve divided by sample size $N$. With a variance estimate, we can create Wald-Type 95\% confidence intervals and test the null hypothesis. Additionally, building on the principle of empirical efficiency maximization \citep{Rubin2008}, we can use the estimated influence curve-squared as our measure of performance in APS, as outlined below \citep{Balzer2016DataAdapt}.

For the population parameter $\psi^p(a)$, the  estimated influence curve for TMLE for observation $O_i$ is given by
\begin{align}
\label{eq:ICpop_notarget}
 \hat{D}^{p}(a;O_i)= \left(\frac{\mathbbm{1}(A_i=a)}{\mathbb{P}(A=a|W_i)}\right)\left(Y_i - \hat{\mathbb{E}}(Y|A,W_i)\right) + \hat{\mathbb{E}}(Y|A=a,W_i) - \hat{\psi}^{p}(a) 
 \end{align}
 when the propensity score $\mathbb{P}(A=a|W)$ is treated as known (i.e., no targeting is done) and
 \begin{equation}
\label{eq:ICpop}
 \hat{D}^{p*}(a;O_i)= \left(\frac{\mathbbm{1}(A_i=a)}{\hat{\mathbb{P}}(A=a|W_i)}\right)\left(Y_i - \hat{\mathbb{E}}^*(Y|A,W_i)\right) + \hat{\mathbb{E}}^*(Y|A=a,W_i) - \hat{\psi}^{p*}(a)
\end{equation}
when we have estimated the propensity score $\hat{\mathbb{P}}(A=a|W)$ and obtained a targeted estimator.
   The latter provides a conservative approximation (pp. 96  of \cite{MarkBook}).
  As detailed in \cite{Balzer2015Adaptive, Balzer2016SATE}, the influence curve of TMLE for both the conditional parameter $\psi^c(a)$  and the sample parameter $\psi^s(a)$  is conservatively approximated by
  \begin{align}
\label{eq:ICsample_notarget}
 \hat{D}^{c}(a;O_i)  = \hat{D}^{s}(a;O_i)= \left(\frac{\mathbbm{1}(A_i=a)}{\mathbb{P}(A=a|W_i)}\right)\left(Y_i - \hat{\mathbb{E}}(Y|A,W_i)\right) 
 \end{align}
  when the propensity score $\mathbb{P}(A=a|W)$ is treated as known (i.e., no targeting is done) and
\begin{equation}
\label{eq:ICsample}
 \hat{D}^{c*}(a;O_i)= \hat{D}^{s*}(a;O_i)= \left(\frac{\mathbbm{1}(A_i=a)}{\hat{\mathbb{P}}(A=a|W_i)}\right)\left(Y_i - \hat{\mathbb{E}}^*(Y|A,W_i)\right) 
 \end{equation}
when we have estimated the propensity score $\hat{\mathbb{P}}(A=a|W)$ and obtained a targeted estimator.

By applying the Delta Method, we can obtain inference for any contrast of treatment-specific means (e.g., the difference, ratio, or odds ratio).
For example, the estimated influence curve for TMLE for the PATE=$\psi^p(1)-\psi^p(0)$ is given by 
$ \hat{D}^{p*}(1;O) -  \hat{D}^{p*}(0;O)$. For relative effects, it is easier to obtain inference on logarithmic scale and then transform back \citep{Moore2009}. For example,  the estimated influence curve for TMLE of the sample risk ratio  on the logarithmic scale $log\{\psi^s(1)/\psi^s(0)\}$ is given by $\hat{D}^{s*}(1;O)/\hat{\psi}^{s*}(1) - \hat{D}^{s*}(0;O)/\hat{\psi}^{s*}(0)$.

\subsection*{A.3. Maximizing empirical efficiency with Adaptive Prespecification (APS)}

The goal of APS is to have a fully automated and data-adaptive approach to select, from a prespecified set, the candidate estimation approach that maximizes precision. A schematic of APS to select the optimal TMLE is given in Figure 1 of the main text. We provide additional mathematical details here.

We consider $V$-fold cross-validation, and let $K_v$ denote the set of indices for the observations in fold $v$. Additionally, let superscript $-v$ denote estimators fit on data excluding observations in fold $v$. First, we select the candidate TMLE to minimize the cross-validated variance estimate when treating the propensity score as known:
\[
 \mbox{CV-Risk} = \frac{1}{V} \sum_{v=1}^V \left[ \frac{1}{n_v} \sum_{k \in K_v}   \big\{ \hat{D}^{-v}(O_k)  \big\} ^2 \right]
\]
 where $n_v=|K_v|$ denotes the number of observations in fold $v$ and $\hat{D}^{-v}(O_k)$ denotes the cross-validated estimate of influence curve for the TMLE without targeting (Eqs.~\ref{eq:ICpop_notarget} and \ref{eq:ICsample_notarget}). For example, for the PATE=$\psi^p(1)-\psi^p(0)$, we would chose the candidate estimator of the outcome regression that minimized  $\frac{1}{V} \sum_{v=1}^V \frac{1}{n_v} \sum_{k \in K_v}  \big\{ \hat{D}^{p,-v}(1,O_k) - \hat{D}^{p,-v}(0,O_k) \big\} ^2 $.
For  the  SATE=$\psi^s(1)-\psi^s(0)$ in a two-armed trial with equal allocation probability $\mathbb{P}(A=1)=0.5$, this corresponds to selecting the candidate outcome regression estimator that  minimizes the L2 loss 
\citep{Balzer2016SATE}

Second, given the selected outcome regression estimator, we select the candidate TMLE, which further minimizes the cross-validated variance estimate after targeting based on the candidate estimator of the propensity score:
\[
 \mbox{CV-Risk} = \frac{1}{V} \sum_{v=1}^V \left[ \frac{1}{n_v} \sum_{k \in K_v}   \big\{ \hat{D}^{*,-v}(O_k)  \big\} ^2 \right]
\]
 where $\hat{D}^{*,-v}(O_k)$ denotes the cross-validated estimate of influence curve for the TMLE \emph{after} targeting (Eqs.~\ref{eq:ICpop} and \ref{eq:ICsample}). For example, for the PATE=$\psi^p(1)-\psi^p(0)$, we would chose the candidate TMLE minimizing \\ $\frac{1}{V} \sum_{v=1}^V \frac{1}{n_v} \sum_{k \in K_v}  \big\{ \hat{D}^{p*,-v}(1,O_k) - \hat{D}^{p*,-v}(0,O_k) \big\} ^2 $. Importantly, the unadjusted estimator of the known propensity score is always included. Therefore, if adjusting during propensity score estimation does not further improve precision, the algorithm defaults to the TMLE only adjusting with the optimal estimator for the outcome regression. Equally important, the unadjusted estimator of the outcome regression is always included. Therefore, if adjusting when fitting the outcome regression does not improve precision, the algorithm defaults to unadjusted effect estimator.

\subsection*{A.4. Alternative designs}

Throughout, we have focused on two-armed trials with simple randomization.
For discussion of TMLE applied in trials with covariate-adaptive randomization schemes, such as pair-matching and stratification, we refer the reader to  \cite{Balzer2015Adaptive, Balzer2016SATE, wang2021}. For extensions to cluster randomized trials, including inference for effects defined at the cluster- or individual-level and inference with arm-specific dependence structures, we refer the reader to \cite{Benitez2021, NugentThesis2022}.

\section*{Appendix B. More details on the simulation}

\subsection*{B.1. Data generating processes for the continuous outcome simulations}

To study performance with a continuous outcome and on the difference scale, we conducted the following simulation study. For 5000 simulated trials each with 500 participants, we generated  5 measured covariates and 2 unmeasured covariates:
$W_1 \sim Bern(p=0.5)$, $W_2 \sim Bern(p=0.2W_1)$, $W_3 \sim Unif(0,5)$, $W_4=logit^{-1}(-2+W_1+W_2+ Unif(0,2))$, $W_5=1+Binom(3, p=0.3)$, $U_1 \sim Unif(0,.5)$, and $U_{2} \sim Unif(0,1)$, respectively. We then generated the continuous, counterfactual outcomes $Y(a)$ in 3 settings of varying complexity: 
\begin{enumerate}
\item``Linear'': $Y(a) = 90 + .07a + .7W_1 + .3W_2 + .1W_3 + .3W_4 + .4W_5 + 0.25aW_1 + 5U_1 + U_2$
\item``Interactive'':  $Y(a) = 150 + .05a + .33W_1 - .25W_2 + .5W_3 - .2W_4 + .05W_5 + .01aW_1 + .02aW_3 + .3aU_1 + 5.8U_1 + U_2$
\item ``Polynomial'':  $Y(a) =  90 + .17a + .33(W_1+W_2+W_3+W_4+W_5) - .2W_1W_3 + .5W_1(.8-.6W_4)W_3 + .25(1-W_1)(-.2+ .15W_4) + 4.7U_1 + U_2$.
\end{enumerate}
We additionally considered a ``Treatment only'' scenario where none of the measured covariates influences the outcome: $Y(a) = 90 + .1a + 3U_1 + U_2$.
As before, we generated the observed treatment $A$ using simple randomization and randomization within strata defined by $\mathbbm{1}(W_1>0)$. Again, we set the observed outcome $Y$ equal to the counterfactual outcome $Y(a)$ when $A=a$.

\subsection*{B.2. Additional simulation results}

In following tables, we provide additional simulation results. Throughout ``DGP'' denotes the data generating process, and ``Design'' refers to whether the trial used simple randomization or stratified randomization. 

In Tables~\ref{Table:BinSelect} and~\ref{Table:ContSelect}, we summarize the selection of candidate estimators for the outcome regression and for the propensity score in the large-trial APS implementation for the 5000 simulated trials when there was an effect. The candidate estimators for the outcome regression were the unadjusted estimator (``Unadjust''), a working  generalized linear model adjusting for a single baseline covariate (``GLM''), main terms regression adjusting for all covariates (``Main''), stepwise regression (``Step''),  stepwise regression with all possible pairwise interactions (``StepInt''), least absolute shrinkage and selection operator (``LASSO''), and   to multivariate adaptive regression splines (``MARS''). For the propensity score, we used a similar set of candidates, but excluded stepwise regression with interactions and MARS. We see that selection is adaptive to the data generating process for both outcome types.

\begin{table}
\caption{Proportion of times (\%) a candidate algorithm was selected for estimation of the outcome regression (``OutReg'') and the propensity score (``PScore'') in TMLE with the large-trial APS in the \textbf{binary outcome} setting when there was an effect. Stepwise regression with all possible pairwise interactions and MARS were not considered as candidates for the propensity score.}
\label{Table:BinSelect}
\centering
\begin{tabular}{llllllllll}
  \hline
Target & DGP & Design & Unadj & GLM & Main & Step & StepInt & LASSO & MARS \\ 
  \hline
OutReg & Linear & Simple & 0\% & 0\% & 0.2\% & 0\% & 98.1\% & 0.1\% & 1.6\% \\ 
   & Linear & Stratified & 0\% & 0\% & 0.1\% & 0\% & 98\% & 0.1\% & 1.8\% \\ 
   & Interactive & Simple & 0\% & 0\% & 25.5\% & 0\% & 17.3\% & 56.9\% & 0.3\% \\ 
   & Interactive & Stratified & 0\% & 0\% & 25.3\% & 0\% & 17\% & 57.2\% & 0.4\% \\ 
   & Polynomial & Simple & 0\% & 0\% & 31.7\% & 0\% & 12.1\% & 55.7\% & 0.5\% \\ 
   & Polynomial & Stratified & 0\% & 0\% & 31.3\% & 0\% & 12.1\% & 55.9\% & 0.7\% \\ 
  PScore & Linear & Simple & 34.9\% & 64.3\% & 0.2\% & 0\% & - & 0.6\% & - \\ 
   & Linear & Stratified & 37.3\% & 62.2\% & 0.1\% & 0\% & - & 0.4\% & - \\ 
   & Interactive & Simple & 35\% & 64.2\% & 0.2\% & 0\% & - & 0.6\% & - \\ 
   & Interactive & Stratified & 36\% & 63\% & 0.3\% & 0\% & - & 0.7\% & - \\ 
   & Polynomial & Simple & 40\% & 59.4\% & 0.2\% & 0\% & - & 0.4\% & - \\ 
   & Polynomial & Stratified & 39.9\% & 59.3\% & 0.2\% & 0\% & - & 0.6\% & - \\ 
   \hline
\end{tabular}
\end{table}

\begin{table}
\caption{Proportion of times (\%) a candidate algorithm was selected for estimation of the outcome regression (``OutReg'') and the propensity score (``PScore'') in TMLE with the large-trial APS in the \textbf{continuous outcome} setting when there was an effect. Stepwise regression with all possible pairwise interactions and MARS were not considered as candidates for the propensity score.}
\label{Table:ContSelect}
\centering
\begin{tabular}{llllllllll}
  \hline
Target & DGP & Design & Unadj & GLM & Main & Step & StepInt & LASSO & MARS \\ 
  \hline
OutReg & Linear & Simple & 0\% & 0\% & 6.7\% & 0\% & 0\% & 91.9\% & 1.4\% \\ 
   & Linear & Stratified & 0\% & 0\% & 5.8\% & 0\% & 0\% & 92.8\% & 1.3\% \\ 
   & Interactive & Simple & 0\% & 0.5\% & 9.4\% & 0\% & 0\% & 89.4\% & 0.7\% \\ 
   & Interactive & Stratified & 0\% & 0.6\% & 8.3\% & 0\% & 0\% & 90.4\% & 0.7\% \\ 
   & Polynomial & Simple & 0\% & 0\% & 13\% & 0\% & 0\% & 86.5\% & 0.5\% \\ 
   & Polynomial & Stratified & 0\% & 0\% & 12\% & 0\% & 0\% & 87.6\% & 0.5\% \\ 
  PScore & Linear & Simple & 68.8\% & 31.2\% & 0\% & 0\% & - & 0\% & - \\ 
   & Linear & Stratified & 72.5\% & 27.5\% & 0\% & 0\% & - & 0\% & - \\ 
   & Interactive & Simple & 70.5\% & 29.5\% & 0\% & 0\% & - & 0\% & - \\ 
   & Interactive & Stratified & 73.5\% & 26.5\% & 0\% & 0\% & - & 0\% & - \\ 
   & Polynomial & Simple & 68.2\% & 31.8\% & 0\% & 0\% & - & 0\% & - \\ 
   & Polynomial & Stratified & 71.9\% & 28.1\% & 0\% & 0\% & - & 0\% & - \\ 
   \hline
\end{tabular}
\end{table}

 In Tables~\ref{Table:BinEffectNULL} and~\ref{Table:ContEffectNull}, we provide simulation results when there is no effect (i.e., under the null).
``Cover.'' denotes the 95\% confidence interval coverage;  ``Type-I'' denotes the proportion of trials where the true null hypothesis was rejected; ```MSE'' denotes the mean squared error; ``Var.'' denotes the variance of the point estimates, and ``Rel.Eff.'' denotes relative efficiency, approximated by the ratio of the MSE of a given estimator to that of the unadjusted estimator. ``Unadjusted'' refers to the unadjusted effect estimator, ``Static'' to forced adjustment for $W_1$ in the outcome regression, ``Small APS'' to TMLE with the small-trial implementation of APS, and ``Large APS'' to TMLE with the large-trial implementation of APS.

\begin{table}
\caption{Simulation results for the \textbf{binary outcome under the null}; the true value of the sample risk ratio is 1. ``Txt only'' refers to the setting when there are no prognostic covariates.}
\label{Table:BinEffectNULL}
\centering
\begin{tabular}{l l l llllll}
  \hline
DGP & Design & Estimator & Cover. & Type-I & MSE & Bias & Var. & Rel.Eff. \\ 
  \hline
Txt only & Simple & Unadjusted & 0.952 & 0.048 & 0.003 & 0.001 & 0.003 & 1.000 \\ 
   &  & Static & 0.951 & 0.049 & 0.003 & 0.001 & 0.003 & 1.001 \\ 
   &  & Small APS & 0.952 & 0.048 & 0.003 & 0.001 & 0.003 & 0.988 \\ 
   &  & Large APS & 0.953 & 0.047 & 0.003 & 0.001 & 0.003 & 0.983 \\ 
   & Stratified & Unadjusted & 0.951 & 0.049 & 0.003 & 0.001 & 0.003 & 1.000 \\ 
   &  & Static & 0.950 & 0.050 & 0.003 & 0.001 & 0.003 & 1.000 \\ 
   &  & Small APS & 0.951 & 0.049 & 0.003 & 0.001 & 0.003 & 0.988 \\ 
   &  & Large APS & 0.952 & 0.048 & 0.003 & 0.001 & 0.003 & 0.988 \\ 
  Linear & Simple & Unadjusted & 0.953 & 0.047 & 0.006 & 0.003 & 0.006 & 1.000 \\ 
   &  & Static & 0.952 & 0.048 & 0.005 & 0.003 & 0.005 & 0.900 \\ 
   &  & Small APS & 0.963 & 0.037 & 0.005 & 0.003 & 0.005 & 0.783 \\ 
   &  & Large APS & 0.950 & 0.050 & 0.004 & 0.002 & 0.004 & 0.659 \\ 
   & Stratified & Unadjusted & 0.956 & 0.044 & 0.006 & 0.002 & 0.006 & 1.000 \\ 
   &  & Static & 0.949 & 0.051 & 0.005 & 0.002 & 0.005 & 0.958 \\ 
   &  & Small APS & 0.962 & 0.038 & 0.005 & 0.002 & 0.005 & 0.824 \\ 
   &  & Large APS & 0.949 & 0.051 & 0.004 & 0.001 & 0.004 & 0.698 \\ 
  Interactive & Simple & Unadjusted & 0.954 & 0.046 & 0.003 & 0.002 & 0.003 & 1.000 \\ 
   &  & Static & 0.953 & 0.047 & 0.003 & 0.001 & 0.003 & 0.898 \\ 
   &  & Small APS & 0.961 & 0.039 & 0.002 & 0.001 & 0.002 & 0.793 \\ 
   &  & Large APS & 0.945 & 0.055 & 0.002 & 0.001 & 0.002 & 0.677 \\ 
   & Stratified & Unadjusted & 0.955 & 0.045 & 0.003 & 0.001 & 0.003 & 1.000 \\ 
   &  & Static & 0.946 & 0.054 & 0.003 & 0.001 & 0.003 & 0.962 \\ 
   &  & Small APS & 0.959 & 0.041 & 0.002 & 0.001 & 0.002 & 0.822 \\ 
   &  & Large APS & 0.939 & 0.061 & 0.002 & 0.001 & 0.002 & 0.730 \\ 
  Polynomial & Simple & Unadjusted & 0.951 & 0.049 & 0.005 & 0.004 & 0.005 & 1.000 \\ 
   &  & Static & 0.945 & 0.055 & 0.004 & 0.003 & 0.004 & 0.838 \\ 
   &  & Small APS & 0.954 & 0.046 & 0.004 & 0.003 & 0.004 & 0.771 \\ 
   &  & Large APS & 0.943 & 0.057 & 0.003 & 0.002 & 0.003 & 0.633 \\ 
   & Stratified & Unadjusted & 0.964 & 0.036 & 0.004 & 0.003 & 0.004 & 1.000 \\ 
   &  & Static & 0.951 & 0.049 & 0.004 & 0.003 & 0.004 & 0.927 \\ 
   &  & Small APS & 0.959 & 0.041 & 0.003 & 0.002 & 0.003 & 0.847 \\ 
   &  & Large APS & 0.947 & 0.053 & 0.003 & 0.002 & 0.003 & 0.733 \\ 
   \hline
\end{tabular}
\end{table}

\begin{table}
\caption{Estimator performance with the \textbf{continuous outcome under the null}; the true value of the SATE is 0. ``Txt only'' refers to the setting when there are no prognostic covariates.}
\label{Table:ContEffectNull}
\centering
\begin{tabular}{l l l llllll}
  \hline
DGP & Design & Estimator & Cover. & Type-I & MSE & Bias & Var. & Rel.Eff. \\ 
\hline
Txt only & Simple & Unadjusted & 0.950 & 0.050 & 0.002 & -0.001 & 0.002 & 1.000 \\ 
   &  & Static & 0.949 & 0.051 & 0.002 & -0.001 & 0.002 & 1.003 \\ 
   &  & Small APS & 0.949 & 0.051 & 0.002 & -0.001 & 0.002 & 1.006 \\ 
   &  & Large APS & 0.950 & 0.050 & 0.002 & -0.001 & 0.002 & 1.005 \\ 
   & Stratified & Unadjusted & 0.945 & 0.055 & 0.002 & 0.000 & 0.002 & 1.000 \\ 
   &  & Static & 0.945 & 0.055 & 0.002 & 0.000 & 0.002 & 1.000 \\ 
   &  & Small APS & 0.946 & 0.054 & 0.002 & 0.000 & 0.002 & 1.003 \\ 
   &  & Large APS & 0.945 & 0.055 & 0.002 & 0.000 & 0.002 & 1.005 \\ 
  Linear & Simple & Unadjusted & 0.953 & 0.047 & 0.007 & -0.001 & 0.007 & 1.000 \\ 
   &  & Static & 0.947 & 0.053 & 0.006 & -0.002 & 0.006 & 0.836 \\ 
   &  & Small APS & 0.948 & 0.052 & 0.006 & -0.002 & 0.006 & 0.825 \\ 
   &  & Large APS & 0.947 & 0.053 & 0.005 & -0.001 & 0.005 & 0.664 \\ 
   & Stratified & Unadjusted & 0.966 & 0.034 & 0.006 & 0.001 & 0.006 & 1.000 \\ 
   &  & Static & 0.945 & 0.055 & 0.006 & 0.001 & 0.006 & 1.000 \\ 
   &  & Small APS & 0.948 & 0.052 & 0.006 & 0.001 & 0.006 & 0.969 \\ 
   &  & Large APS & 0.944 & 0.056 & 0.005 & 0.001 & 0.005 & 0.799 \\ 
  Interactive & Simple & Unadjusted & 0.948 & 0.052 & 0.011 & -0.001 & 0.011 & 1.000 \\ 
   &  & Static & 0.946 & 0.054 & 0.011 & -0.001 & 0.011 & 0.990 \\ 
   &  & Small APS & 0.948 & 0.052 & 0.007 & -0.001 & 0.007 & 0.611 \\ 
   &  & Large APS & 0.946 & 0.054 & 0.006 & -0.001 & 0.006 & 0.585 \\ 
   & Stratified & Unadjusted & 0.942 & 0.058 & 0.011 & 0.003 & 0.011 & 1.000 \\ 
   &  & Static & 0.939 & 0.061 & 0.011 & 0.003 & 0.011 & 1.000 \\ 
   &  & Small APS & 0.947 & 0.053 & 0.007 & 0.001 & 0.007 & 0.601 \\ 
   &  & Large APS & 0.943 & 0.057 & 0.006 & 0.001 & 0.006 & 0.583 \\ 
  Polynomial & Simple & Unadjusted & 0.948 & 0.052 & 0.008 & -0.001 & 0.008 & 1.000 \\ 
   &  & Static & 0.943 & 0.057 & 0.007 & -0.001 & 0.007 & 0.938 \\ 
   &  & Small APS & 0.952 & 0.048 & 0.006 & -0.002 & 0.006 & 0.729 \\ 
   &  & Large APS & 0.948 & 0.052 & 0.004 & -0.001 & 0.004 & 0.585 \\ 
   & Stratified & Unadjusted & 0.948 & 0.052 & 0.007 & 0.002 & 0.007 & 1.000 \\ 
   &  & Static & 0.940 & 0.060 & 0.007 & 0.002 & 0.007 & 1.000 \\ 
   &  & Small APS & 0.957 & 0.043 & 0.005 & 0.001 & 0.005 & 0.712 \\ 
   &  & Large APS & 0.942 & 0.058 & 0.005 & 0.001 & 0.005 & 0.620 \\ 
   \hline
\end{tabular}
\end{table}

\section*{Appendix C. Additional information and results for the real data application}

In Table~\ref{Table:RealChar}, we provide the baseline characteristics, used as candidate adjustment variables, in the application to  ACTG 175 Study \citep{ACTG175}, whose data are publicly available in the \texttt{speff2trial} \texttt{R} package \citep{speff2trial}. While there was an imbalance in the number randomized to the intervention $(N=1587)$ and to the control $(N=526$), the baseline covariates were well-balanced between arms.

\begin{table}[ht]
\caption{Baseline characteristics and candidate adjustment variables in the real data application to the ACTG 175 Study, by arm and overall \citep{ACTG175}. Continuous variables are shown in median [Q1, Q3], and binary variables are shown in N (\%).}
\label{Table:RealChar}
\centering
\begin{tabular}{llll}
  \hline
 & Intervention & Control & Overall \\ 
  & N=1587 & N=526 & N=2113 \\ 
  \hline
  Age (years) & 34 [29,40] & 34 [29,40] & 34 [29,40] \\ 
  Aged 18-29 years & 399 (25\%) & 137 (26\%) & 536 (25\%) \\ 
  Male & 1319 (83\%) & 427 (81\%) & 1746 (83\%) \\ 
  Non-white race & 456 (29\%) & 154 (29\%) & 610 (29\%) \\ 
  Weight (kg) & 74 [67,82] & 75 [68,84] & 74 [67,83] \\ 
  Has hemophilia & 118 (7\%) & 37 (7\%) & 155 (7\%) \\ 
  Karnofsky score (scale 0-100) & 100 [90,100] & 100 [90,100] & 100 [90,100] \\ 
  Symptomatic & 279 (18\%) & 88 (17\%) & 367 (17\%) \\ 
  ART experienced & 926 (58\%) & 307 (58\%) & 1233 (58\%) \\ 
  Time on ART (days) & 139 [0,739] & 138 [0,731] & 139 [0,735] \\ 
  Started ART 1-52wks prior & 312 (20\%) & 96 (18\%) & 408 (19\%) \\ 
  Non-zidovudine prior to baseline & 31 (2\%) & 16 (3\%) & 47 (2\%) \\ 
  Baseline CD4 count (cells/mm$^3$) & 339 [260,423] & 346 [271,422] & 340 [263,423] \\ 
  Baseline CD4$>$350 & 724 (46\%) & 252 (48\%) & 976 (46\%) \\ 
  Baseline CD8 count (cells/mm$^3$) & 897 [655,1212] & 880 [656,1190] & 894 [655,1210] \\ 
  Baseline CD48$>$350 & 1540 (97\%) & 520 (99\%) & 2060 (97\%) \\ 
   \hline
\end{tabular}
\end{table}

In  Tables~\ref{Table:WebRealCont} and~\ref{Table:WebRealBin}, we provide additional results for the effect on average CD4 count (continuous outcome) and the probability of having a CD4 count$>$350 c/mm$^3$ (binary outcome), overall and within subgroups defined by age (18-29 years and 30+ years) and gender. A coding vignette to reproduce these and additional results is available at \url{https://github.com/LauraBalzer/AdaptivePrespec}.

``Small TMLE'' and ``Large TMLE'' refer to using APS  to only select the outcome regression estimator in the small-trial and large-trial implementation, respectively. ``Small CTMLE'' and ``Large CTMLE'' refer to using APS for selection of the outcome regression estimator and collaborative selection of the known propensity score estimator in the small-trial and large-trial implementation, respectively. ``Out.Reg.'' and ``PScore'' refer to the fixed or adaptively selected estimator of the outcome regression and propensity score, respectively. For estimation of  the outcome regression and the propensity score, the large-trial APS considered the unadjusted estimator, adjustment for a single covariate a working generalized linear model (``GLM''), a main terms regression adjusting for all covariates (``Main''), stepwise regression, LASSO, MARS, and MARS after screening based on pairwise correlations. As before, the small-trial implementation was limited to the unadjusted estimator and adjustment for a single covariate a working generalized linear model.

\begin{table}[ht]
\caption{Additional results from the real data application for the additive effect on the \textbf{continuous outcome}: CD4 count at week 20.  ``Younger'' refers to age 18-29 years, while ``older'' refers to age 30+ years.}
\label{Table:WebRealCont}
\centering
\begin{tabular}{llllll}
  \hline
 & Estimator & Effect (95\%CI) & Rel.Var. & Out.Reg. & PScore \\ 
 \hline
Overall  & Unadjusted & 46.4 (33.0, 59.7) & 1.000 & Unadj. & Unadj. \\ 
 (N=2113)  & Static & 46.8 (33.5, 60.0) & 0.991 & Fixed & Fixed \\ 
   & Small TMLE & 48.4 (37.9, 58.9) & 0.621 & GLM & Unadj. \\ 
   & Small CTMLE & 48.5 (38.0, 59.0) & 0.617 & GLM & GLM \\ 
   & Large TMLE & 47.7 (37.9, 57.6) & 0.546 & MARS & Unadj. \\ 
   & Large CTMLE & 47.8 (38.0, 57.6) & 0.542 & MARS & GLM \\ 
  Older women & Unadjusted & 53.6 (18.4, 88.7) & 1.000 & Unadj. & Unadj. \\ 
  (N=258) & Static & 53.3 (18.5, 88.1) & 0.979 & Fixed & Fixed \\ 
   & Small TMLE & 62.5 (32.5, 92.4) & 0.723 & GLM & Unadj. \\ 
   & Small CTMLE & 64.3 (35.3, 93.4) & 0.682 & GLM & GLM \\ 
   & Large TMLE & 62.5 (32.5, 92.4) & 0.723 & GLM & Unadj. \\ 
   & Large CTMLE & 64.3 (35.3, 93.4) & 0.682 & GLM & GLM \\ 
  Younger women & Unadjusted & -13.6 (-81.4, 54.2) & 1.000 & Unadj. & Unadj. \\ 
  (N=109) & Static & -14.0 (-81.8, 53.9) & 1.002 & Fixed & Fixed \\ 
   & Small TMLE & 29.6 (-15.8, 75.1) & 0.449 & GLM & Unadj. \\ 
   & Small CTMLE & 37.3 (-4.9, 79.4) & 0.386 & GLM & GLM \\ 
   & Large TMLE & 29.6 (-15.8, 75.1) & 0.449 & GLM & Unadj. \\ 
   & Large CTMLE & 37.3 (-4.9, 79.4) & 0.386 & GLM & GLM \\ 
  Older men & Unadjusted & 50.8 (34.3, 67.4) & 1.000 & Unadj. & Unadj. \\ 
  (N=1319) & Static & 50.8 (34.2, 67.3) & 0.999 & Fixed & Fixed \\ 
   & Small TMLE & 50.2 (37.0, 63.4) & 0.637 & GLM & Unadj. \\ 
   & Small CTMLE & 50.3 (37.1, 63.4) & 0.636 & GLM & GLM \\ 
   & Large TMLE & 48.9 (36.2, 61.6) & 0.591 & LASSO & Unadj. \\ 
   & Large CTMLE & 48.9 (36.2, 61.6) & 0.591 & LASSO & GLM \\ 
  Younger men & Unadjusted & 47.4 (17.7, 77.1) & 1.000 & Unadj. & Unadj. \\ 
  (N=427) & Static & 46.2 (16.6, 75.8) & 0.994 & Fixed & Fixed \\ 
   & Small TMLE & 42.0 (19.1, 65.0) & 0.599 & GLM & Unadj. \\ 
   & Small CTMLE & 41.2 (18.3, 64.2) & 0.598 & GLM & GLM \\ 
   & Large TMLE & 45.6 (23.9, 67.4) & 0.538 & Main & Unadj. \\ 
   & Large CTMLE & 45.0 (23.3, 66.7) & 0.534 & Main & GLM \\ 
   \hline
\end{tabular}
\end{table}

\begin{table}[ht]
\caption{Additional results from the real data application for the relative effect on the \textbf{binary outcome}: having a CD4 count$>$350 c/mm$^3$ at week 20. Young'' refers to age 18-29 years, while ``old'' refers to age 30+ years.}
\label{Table:WebRealBin}
\centering
\begin{tabular}{llllll}
  \hline
 & Estimator & Effect (95\%CI) & Rel.Var. & Out.Reg. & PScore \\ 
  \hline
Overall & Unadjusted & 1.23 (1.10, 1.37) & 1.000 & Unadj. & Unadj. \\ 
  (N=2113) & Static & 1.23 (1.11, 1.37) & 1.001 & Fixed & Fixed \\ 
   & Small TMLE & 1.26 (1.15, 1.38) & 0.707 & GLM & Unadj. \\ 
   & Small CTMLE & 1.26 (1.15, 1.38) & 0.702 & GLM & GLM \\ 
   & Large TMLE & 1.25 (1.15, 1.37) & 0.678 & LASSO & Unadj. \\ 
   & Large CTMLE & 1.26 (1.15, 1.37) & 0.672 & LASSO & GLM \\ 
  Older women & Unadjusted & 1.21 (0.89, 1.65) & 1.000 & Unadj. & Unadj. \\ 
  (N=258) & Static & 1.21 (0.89, 1.65) & 0.984 & Fixed & Fixed \\ 
   & Small TMLE & 1.28 (0.96, 1.71) & 0.877 & GLM & Unadj. \\ 
   & Small CTMLE & 1.27 (0.96, 1.68) & 0.847 & GLM & GLM \\ 
   & Large TMLE & 1.28 (0.96, 1.71) & 0.877 & GLM & Unadj. \\ 
   & Large CTMLE & 1.27 (0.96, 1.69) & 0.841 & GLM & GLM \\ 
  Younger women & Unadjusted & 1.05 (0.71, 1.55) & 1.000 & Unadj. & Unadj. \\ 
  (N=109) & Static & 1.04 (0.71, 1.53) & 1.000 & Fixed & Fixed \\ 
   & Small TMLE & 1.35 (1.00, 1.83) & 0.606 & GLM & Unadj. \\ 
   & Small CTMLE & 1.35 (1.01, 1.81) & 0.571 & GLM & GLM \\ 
   & Large TMLE & 1.35 (1.00, 1.83) & 0.606 & GLM & Unadj. \\ 
   & Large CTMLE & 1.35 (1.01, 1.81) & 0.571 & GLM & GLM \\ 
  Older men & Unadjusted & 1.25 (1.09, 1.45) & 1.000 & Unadj. & Unadj. \\ 
  (N=1319) & Static & 1.25 (1.09, 1.44) & 1.000 & Fixed & Fixed \\ 
   & Small TMLE & 1.25 (1.11, 1.41) & 0.702 & GLM & Unadj. \\ 
   & Small CTMLE & 1.25 (1.11, 1.41) & 0.702 & GLM & GLM \\ 
   & Large TMLE & 1.25 (1.11, 1.40) & 0.672 & LASSO & Unadj. \\ 
   & Large CTMLE & 1.25 (1.11, 1.40) & 0.669 & LASSO & GLM \\ 
  Younger men & Unadjusted & 1.23 (0.98, 1.56) & 1.000 & Unadj. & Unadj. \\ 
  (N=427) & Static & 1.23 (0.97, 1.55) & 0.988 & Fixed & Fixed \\ 
   & Small TMLE & 1.24 (1.03, 1.49) & 0.616 & GLM & Unadj. \\ 
   & Small CTMLE & 1.24 (1.03, 1.49) & 0.612 & GLM & GLM \\ 
   & Large TMLE & 1.23 (1.03, 1.46) & 0.549 & LASSO & Unadj. \\ 
   & Large CTMLE & 1.22 (1.03, 1.45) & 0.528 & LASSO & GLM \\ 
   \hline
\end{tabular}
\end{table}

 \bibliography{mybiblio.bib}

\end{document}